# Ensemble Simulations of Coronal Mass Ejections in Interplanetary Space with Elliptical Cone Models

Johan Muhamad[1][0000-0001-7214-3754] Tiar Dani[1,2][0009-0002-6691-0683] Muhamad Z. Nurzaman[1][0000-0003-2142-530X] Rasdewita Kesumaningrum[1][0009-0000-9896-9861] Santi Sulistiani[1][0009-0004-7340-6279] Farahhati Mumtahana[1][0009-0008-9204-3038] Gerhana P. Putri[1][0009-0004-0979-7295] Ayu D. Pangestu[1][0009-0000-9451-4659], Ahmad Z. Utama[1]

[1] Research Center for Space, National Research and Innovation Agency (BRIN), Bandung 40135, Indonesia
[2] Department of Computer Science and Electronics, Universitas Gadjah Mada, Yogyakarta 55281, Indonesia
johan.muhamad@brin.go.id

**Abstract.** The estimation of CME arrival time strongly depends on the CME propagation models in interplanetary space and the geometrical aspects of the CME model. We conducted ensemble simulations of CMEs propagation with various elliptical cone shapes to study the relation between the CME speed and the optimum cone shape. We numerically searched for the best elliptical aspect ratio of the elliptical cone for each CME in our CME-ICME pair data. We found that the fast CMEs tend to have a higher elliptical aspect ratio (more circular) than the slower CMEs (flattened). Our results suggest that a fast CME gives a stronger push to all directions, which results in a more circular shape of the leading-edge. We believe that this velocity-dependent behavior is related to the different Lorentz force strengths during the early expansion of a CME.

## 1    Introduction

A non-constant coronal mass ejection (CME) speed model, i.e., the drag-based model (DBM), has been developed to accommodate the acceleration or deceleration of a CME in interplanetary space due to the presence of a drag force [1]. The model has undergone significant development, progressing from the fundamental one-dimensional (1D) model to more sophisticated variants including the 2D circular cone self-similar, 2D flattening cone, and the ensemble version of the 2D flattening cone DBM [2].

DBM is developed considering the aerodynamic influence of the interaction between CME and ambient plasma in the interplanetary medium [3]. CMEs moving faster than the ambient solar wind speed tend to decelerate, while those moving slower than the ambient solar wind are accelerated. By considering this interaction, for 1D DBM, the distance of a CME from the Sun as a function of time, $R_t$, can be calculated as [1]:

$$R_t = \frac{S}{\gamma} ln[S\gamma(v_0 - \omega)t + 1] + \omega t + R_0. \qquad (1)$$

$v_0$ is the initial velocity, $\gamma$ is the drag parameter, $\omega$ is the ambient solar wind speed, $R_0$ is the initial distance of a CME when the drag force is dominant compared to the Lorentz force, $t$ is time, and $S$ can be 1 or -1 depending on the relative speed between $v_0$ and $\omega$.

The 2-D implementations typically require a CME cone model to approximate the evolving shape of the CME front during propagation. In a simple 2D DBM, which leading edge of a CME during its propagation is considered to be concentric, the distances of all elements in the leading-edge curvature and the Sun are same. This equidistance can be calculated by using Equation (1). However, a more realistic representation incorporating an elliptical cone was later introduced to better align with observational data.

An elliptical CME evolution DBM was first developed by [4]. It was then further refined by integrating observational data acquired from STEREO/HI [5]. [6] subsequently conducted ensemble simulations utilizing this method, incorporating a broader range of ensemble parameters. Recently, a new elliptical model was developed by [7] that incorporated frontal deformation.

Our study aims to investigate the relationship between cone ellipticity and CME speed by identifying the optimal elliptical aspect ratio that best aligns with observed CME arrival times. We have developed an elliptical cone DBM with an adjustable aspect ratio for tailored fitting to observations. This study focuses on how CME arrival time depends on the ICME cone shape and initial CME velocity. We expect that this relationship between cone ellipticity and CME initial speed will lead to a simple yet effective model for predicting CME arrival times.

## 2 Data and Methods

### 2.1 Data

A set of 75 CME-ICME pairs selected from the interplanetary CME (ICME) database [8], spanning from 1997-2017, is used in our simulations. We only select CMEs that clearly have associated solar flares, neglecting CMEs from quiescent filament eruptions, multiple CME events, and stealth CMEs. The initial CME space speeds are estimated using the de-projected values provided by [9] for the corresponding dataset. The locations of CME sources are determined by the location of the corresponding active region (AR) positions when the associated flare happened. Additionally, data on initial appearance times, speeds, and central position angles of the CMEs are retrieved from the SOHO/LASCO catalog at CDAW (NASA) data center [10].

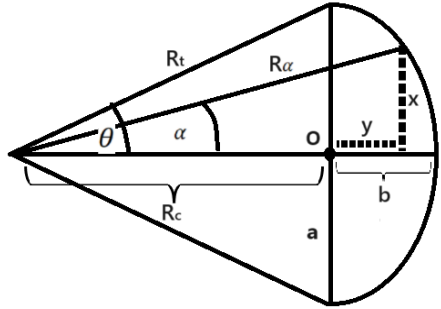

Fig. 1. ICME elliptical cone geometry in the simulations.

## 2.2  Methods

The ICME is modeled as a 2D cone model with the leading edge, which is considered as semi-elliptical shape. The center of this ellipse coincides with the intersection of two lines: one connecting the edges of the ICME legs and another connecting the CME source with the apex (see Fig. 1). The semi-minor axis ($b$) acts as a free parameter, allowing for incremental adjustments to create an adaptable leading-edge that can range from a flattened shape to a semi-circular one. The lengths of the ICME legs ($Rt$) are equally evolving in time according to Equation (1) [1]. The cone half-angle, $\theta$, is determined as a function of the projected CME initial speed based on the statistical relationship proposed by [10]. Considering all the geometrical parameters in Fig. 1, the distance of any element of the elliptical leading-edge from the Sun can be calculated as

$$R_\alpha = \frac{R_c + y}{\cos \alpha}, \qquad (2)$$

where $y$ can be derived by taking advantage of elliptical formula with pre-defined semi major axis ($a$) and $b$,

$$\frac{x^2}{a^2} + \frac{y^2}{b^2} = 1, \qquad (3)$$

and substituting $x$ as,

$$x = \tan \alpha (R_c + y). \qquad (4)$$

Note that our model is not identical to the model by [4], which determined the ellipse center based on the point of tangency between the ellipse and the ICME legs.

Ensemble simulations are employed to determine the optimal aspect ratio ($b/a$) of the ellipse for each CME event within the dataset. This optimization process aimed to achieve the best fit between the simulated arrival time and the corresponding observed arrival time. The simulation models ICME propagation by first calculating ICME legs using Equation (1), and then deriving $R_c$ based on the corresponding cone half angle, $\theta$. Subsequently, it calculates $R_\alpha$ using Equation (2) for all elements from the ICME apex to the ICME legs by incrementally change $\alpha$. Note that the semi major axis, $a$, is constrained by the cone half angle, while $b$ is a free parameter that controls the shape of the cone. This $b/a$ variable is incrementally changed from 0.001 to 1 for flattened to semi-circular cone shape, respectively.

For each CME event, we run the 2D DBM simulation with various aspect ratios and calculate the associated arrival times until we find the one that predicts the arrival time most accurately. The corresponding aspect ratio that can predict the most accurate arrival time is considered as the optimum one. This process is repeated for all CME events.

Since the simulations are based on a pure geometrical approach, actual shapes of the observed CMEs are not fully considered. Therefore, the optimum aspect ratios in the simulations may not accurately reflect the observed CMEs' shapes. This is because the only parameter that constraints the optimization in the simulations is the observed arrival time.

The input parameters obtained from the database are $v_0$, central position angle, and the longitude of the corresponding AR. The drag parameter, $\gamma$, is maintained constant across all simulations at the optimal value of $\gamma=0.2 \times 10^{-7}$ km$^{-1}$ as suggested by [1]. Similarly, a constant ambient solar wind speed of 450 km/s is adopted in the simulations.

## 3     Results and Discussions

Our simulations (Fig. 2) reveal a trend towards a more circular leading-edge (higher $b/a$) for CMEs with higher initial velocities. This relationship strengthens when considering only front-side CMEs (source region longitude < 30°), as depicted by the red circles in Fig. 2. The correlation coefficient ($R$) between the two parameters is significantly higher for front-side CMEs ($R = 0.65$) compared to the entire dataset, including near-limb CMEs, ($R = 0.28$). This may indicate that ICMEs coming from near the disc center can be estimated using a simple elliptical model during its propagation to Earth. Conversely, near-limb CMEs, due to their more complex trajectories, may require a more intricate model for accurate arrival time prediction.

We further investigate the influence of $\gamma$ parameter on the relationship between $b/a$ and $v_0$. Ensemble simulations are conducted using the same dataset but varying $\gamma$ values. The results are presented in Fig. 3, where the left panel depicts the case with $\gamma=0.1 \times 10^{-7}$ km$^{-1}$ and the right panel show the case with $\gamma=0.5 \times 10^{-7}$ km$^{-1}$. The corresponding correlation coefficient (R) between the two parameters for the front-side CMEs are 0.38 and 0.48, respectively.

These results demonstrate that while the overall trend of aspect ratio dependence on initial speed remains relatively consistent, the specific relationship exhibits sensitivity to the chosen $\gamma$ within the simulation. Our preliminary findings suggest that a combination of $\gamma = 0.2 \times 10^{-7}$ km$^{-1}$ and $\omega = 450$ km/s offers a sufficient basis for estimating the appropriate aspect ratio based on initial CME velocity. We realize that the calculation of arrival time in the DBM also depends on the preference of $\omega$, which is constant in our simulations. In reality, the value of $\omega$ can vary for different CME events.

However, for simplicity, we only set the constant value of $\omega$ for all cases during the simulations. This uniform $\omega$ that is applied for all CME events is useful in our simulations to isolate the influence of cone shape to the arrival time prediction, as our main focus in our current study.

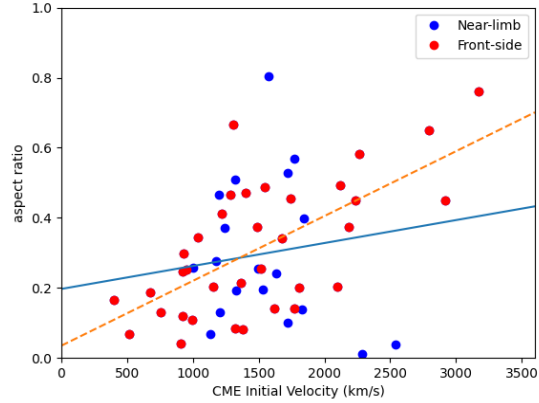

Fig. 2. Plot of the elliptical aspect ratio to the CME initial speed of our simulations for $\gamma = 0.2 \times 10^{-7}$ km$^{-1}$ of the near limb CMEs (blue) and front-side CMEs (red). Blue (orange dashed) line represent the regression line for all (front-side) CMEs.

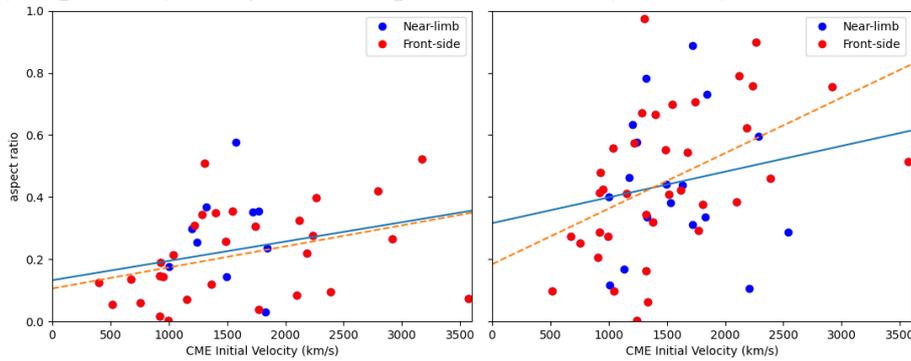

Fig. 3. Plot of the elliptical aspect ratio to the CME initial speed of our simulations for $\gamma=0.1 \times 10^{-7}$ km$^{-1}$ (left) and $\gamma=0.5 \times 10^{-7}$ km$^{-1}$ (right). Color code is the same as in Fig. 2.

## 4   Conclusion

Our findings consistently demonstrate that faster CMEs exhibit a more circular leading-edge on the ICME cone. This suggests a more rapid expansion in all directions for faster CMEs compared to their slower counterparts. We believe that this is related to the strong Lorentz force that act on a flux rope during early expansion of a CME. Our results consistent with previous study that wide CMEs are more impulsive and undergo lateral expansion lower in the corona [12]. However, it should be noted that our study only analyzes limited CME events and relies on assumptions about the constant $\gamma$ and $\omega$. These limitations may not be relevant when analyzing CMEs that have actual drag parameters and ambient solar winds speeds that largely deviate from the assumed values. Ultimately, our results unveil the potential for developing a simplified 2D DBM model that solely requires adjustments to the elliptical aspect ratio based on the CME's initial speed for CME arrival time estimation. A more extensive study using the 2D DBM is currently underway to investigate the effects of different drag parameters and ambient solar wind conditions on CME propagation, while also considering the actual shape of CMEs. The findings of this study will be published in the near future.

## 5   Acknowledgement